\pdfoutput=1
\documentclass[traditabstract]{aa} 
\usepackage{savesymbol}
\usepackage{amsmath}
\savesymbol{iint}
\restoresymbol{TXF}{iint}
\usepackage{txfonts}
\usepackage{float}
\usepackage{placeins}
\usepackage{lscape}
\usepackage{natbib}
\usepackage{ulem}
\newcommand{\fett}[1]{{#1}}
\bibpunct{(}{)}{;}{a}{}{,}
\usepackage{graphicx}
\usepackage{txfonts}
\begin{document}

    \title{Hanny's Voorwerp } 
\subtitle{Evidence of AGN activity and a nuclear starburst in the central regions of IC~2497 }

   \author{H. Rampadarath\thanks{\email{hayden.rampadarath@manchester.ac.uk}} 
          \inst{1,3,4}
          \and
          M.A. Garrett\thanks{\email{garrett@astron.nl}} 
          \inst{2,3,5}
          \and
          G. I. G. J\'ozsa 
           \inst{2}
           \and
          T. Muxlow 
          \inst{4}
           \and
           T. A. Oosterloo
           \inst{2,6}
           \and
           Z. Paragi 
           \inst{1,7}
           \and
           R. Beswick
           \inst{4}
           \and
             H. van Arkel
              \inst{2}
              \and
            W. C. Keel
              \inst{8}
              \and
            K. Schawinski
              \inst{9,10}
           }
   \institute{Joint Institute for VLBI in Europe (JIVE), Postbus 2, 7990 AA Dwingeloo, The Netherlands              
              \and
          Netherlands Institute for Radio Astronomy (ASTRON), Postbus 2, 7990 AA Dwingeloo, The Netherlands
         \and
          Leiden Observatory, Leiden University, P.O. Box 9513, 2300RA Leiden, The Netherlands 
         \and
         Jodrell Bank Centre for Astrophysics, School of Physics and Astronomy, Univ. Manchester, Alan Turing Building,               		Oxford Road, Manchester, M13 9PL, United Kingdom. 
         \and
           Centre for Supercomputing, Swinburne University of Technology, Mail number H39, P.O. Box 218, Hawthorn, 
           Victoria 3122, Australia
           \and
           Kapteyn Astronomical Institute, Univ. Groningen, Postbus 800, 9700 AV Groningen, The Netherlands
           \and
			MTA Research Group for Physical Geodesy and Geodynamics, P.O. Box 91, H-1521 Budapest, Hungary
			\and 
			Univ. Alabama, Dept. Physics \& Astronomy, Box 870324, University of Alabama, Tuscaloosa, AL 35487-0324, 				USA
			\and
			Univ. Yale, Dept. Physics, J.W. Gibbs Laboratory, 260 Whitney Avenue, Yale University, New Haven, CT 06511, 			USA
			\and 
			Einstein Fellow            
             }

   \date{Received; accepted }

  \abstract { We present high- and intermediate resolution radio
  observations of the central region in the spiral galaxy IC~2497, performed using
  the European VLBI Network (EVN) at 18~cm, and the
  Multi-Element Radio Linked Interferometer Network (MERLIN) at 18~cm
  and 6~cm.  We detect two compact radio sources, with brightness
  temperatures above 10$^5$ K, suggesting that they are related to AGN
  activity. We show that the total 18~cm radio emission from the
  galaxy is dominated neither by these compact sources nor
  large-scale emission, but extended emission confined within a sub-kpc
  central region. IC~2497 therefore appears as a typical
  luminous infrared galaxy that exhibits a nuclear starburst with a massive star
  formation rate ($M > 5M_{\odot}$) of 12.4~M$_{\odot}$/yr. These results are
  in line with the hypothesis that the ionisation nebula ``Hanny's
  Voorwerp'' at a distance of $\sim 15-25$ kpc from the galaxy is ionised
  by the radiation cone of the AGN.}


\maketitle

\section{Introduction}
Hanny's Voorwerp (SDSS J094103.80+344334.2) \footnote{'Voorwerp' is
Dutch for object} is an irregular gas cloud located $\sim 25\,\rm kpc$ to
the southeast of the massive disk galaxy IC~2497
\citep{Lintottetal09, Joshetal09}. In the optical, the Voorwerp's
appearance is dominated by [{O}~{III}] emission lines, and its spectrum
shows strong line emission, with high-ionisation lines (He II, [Ne V])
co-extensive with the continuum \citep{Lintottetal09}. Paradoxically,
there is no evidence of an ionising source in the
immediate proximity of this nebulosity. Its quiescent kinematics,
derived from optical spectra, imply that photoionisation is the
predominant ionisation process, rather than ionisation via shocks
\citep{Lintottetal09, Joshetal09}.
The original explanation of this phenomenon was provided by
\citet{Lintottetal09}, who argued that Hanny's Voorwerp may be the
first example of a quasar light echo. It is
proposed that around $10^{5}$ years ago, IC~2497 underwent a
significant outburst with its central luminosity approaching
quasar-like levels before decreasing to current, lower-levels of
activity. \fett{In this scenario}, observations of the Voorwerp today
therefore represent a snapshot of this extreme quasar outburst as it
was $10^{5}$ years ago.

Radio observations using the Westerbork Synthesis Telescope (WSRT) and
the EVN (using the e-VLBI technique) \citep{Joshetal09} have detected
a radio continuum source at the central position of
IC~2497, and \fett{weak, large-scale} emission pointing in the
direction of Hanny's Voorwerp. In addition, neutral hydrogen,
\fett{presumably debris from a past interaction,} is detected around
the galaxy. The Voorwerp is probably part of this large surrounding
gas reservoir. HI is also detected in absorption towards the central
radio core. Obscuring material in the direction of the core
of IC~2497 is clearly present, while the extended continuum implies that a large-scale radio jet is present.
These radio observations support an alternative to the light-echo
scenario, namely that IC~2497 contains an obscured active galactic
nucleus (AGN) with
\fett{a weak, large-scale radio jet} 
pointing in the direction of Hanny's Voorwerp, perpendicular to the
major axis of the galaxy \citep{Joshetal09}. 
Hence, another interpretation of Hanny's Voorwerp is that it is a rare
and spectacular example of \fett{ongoing} AGN feedback in which
\fett{an obscured AGN at the centre of a} relatively nearby galaxy
is observed ionising the surrounding IGM \citep{Joshetal09}.

\fett{
In this Letter, we present new radio continuum observations of the
galaxy IC~2497 with the European VLBI Network (EVN) using the e-VLBI
technique at 18 cm, and with the Multi-Element Radio Linked
Interferometer Network (MERLIN) at 18~cm and 6~cm. The complementary
observations, providing on one hand a higher sensitivity at high
resolution, and on the other hand probing the ISM on intermediate
scales, permit us to investigate the AGN hypothesis and map
the central radio emission, which is unresolved in WSRT- and VLBI
observations. We argue that our observations confirm the former
detection of an AGN at the centre of IC~2497 and indicate the presence
of a strong, central starburst in IC~2497, providing support to
the hypothesis that the AGN activity in IC~2497 is obscured towards
the observer.}

\section{Observations and data reduction}
\label{sect2}
\fett{
In the following discourse, we describe in detail the observations with the EVN
and MERLIN, and the subsequent data analysis. A summary of the
properties of the derived maps is given in Table 1. }

\subsection{e-VLBI 18~cm observations}
\begin{table}
\caption{Radio observations of IC~2497}

\begin{tabular}{cccccc}
\hline  
 Epoch & Array & ${\rm \lambda}$ & $\sigma_{{\rm \lambda}}$ & Beam size & P.A.\\ 
 & & (cm) & $({\rm mJy}\,{\rm beam}^{-1})$ & ($mas^{2}$) & ($^\circ$) \\
\hline  
2009.092& MERLIN & 18 & 0.037 & 178 $\times$ 165 & 19.5 \\ 
 
2009.226 & MERLIN & 6 & 0.070 & 100 $\times$ 100 & 0.0 \\ 

2009.387 & EVN & 18 & 0.015 & 45 $\times$ 27 & 7.7 \\ 
\hline  
\end{tabular} 
\end{table}
\medskip
On 19 May 2009 $\&$ 20 May 2009, IC2497 was observed at
18~cm with the EVN using the e-VLBI technique. The experiment was performed using the Westerbork,
Medicina, Onsala 25-m, Torun, Effelsberg, Jodrell Bank (Lovell),
Cambridge, and Knockin telescopes.

The data were transported to the correlator at the Joint Institute for
VLBI in Europe (JIVE) in real-time using the User Datagram Protocol
(UDP). The radio telescopes were physically connected to JIVE via the
National Research and Educational Networks (NRENs) and the
pan-European research network G\'{E}ANT2 via the Dutch national
research network SURFnet \citep{garrett04, Szomoru2008}. At the time
of our observations, a sustainable data rate of 512 Mbps was
achieved. For further details we refer to \cite{Szomoru2008}.

The target was phase-referenced to J0945+3534 (\cite{VCS1},
RA=09:45:38.121 Dec.=+35:34:55.089, J2000), a phase calibrator located
1.3 degrees away from the target source IC~2497. A
phase-reference cycle time of 7 minutes (2 minutes on the calibrator
and 5 minutes on IC~2497) was employed. The data were composed of 8 x
8 MHz sub-bands (in both left-hand and right-hand polarisations) with
2-bit data sampling at the Nyquist rate, resulting in a total data
rate of 512 Mbps. The correlator accumulation period was 2 seconds and
the total observing time was approximately 10 hours from UT 14:47:36
to 00:54:37,
\fett{with $\sim$7 hours on-source. The 
maximum projected baseline amounts to 8 M$\lambda$
(Lovell-Medicina). The shortest projected baseline (Lovell-Knockin) is
at $\sim$ 0.35 M$\lambda$, hence structures with an angular size of
above $\sim$ 0.6 arcseconds are resolved out (structures with
intensity variations above this scale are not detected).}

The initial data reduction was performed using the National Radio
Astronomical Observatory (NRAO) software package \textit{AIPS}. The
data were edited, amplitude-calibrated and fringe-fitted using
standard techniques. The calibration information was not complete for all the telescopes, and  
we expect the absolute flux density scale to be accurate at the $\pm 20$\% level. The calibration solutions derived from J0945+3534
(including phase and amplitude corrections obtained by hybrid mapping
the source) were applied to the target IC~2497 data. The calibrated UV
data of IC~2497 were Fourier transformed and the CLEAN algorithm \cite{Hogbom74}
was applied using the AIPS task IMAGR. The data were
then phase self-calibrated with corrections derived across the entire 64 MHz 
band in each hand of polarisation and over a solution interval of 10 minutes. 
The final, naturally weighted image is shown in Fig. 1.

\begin{figure}

\centering 
\includegraphics[width=0.45\textwidth]{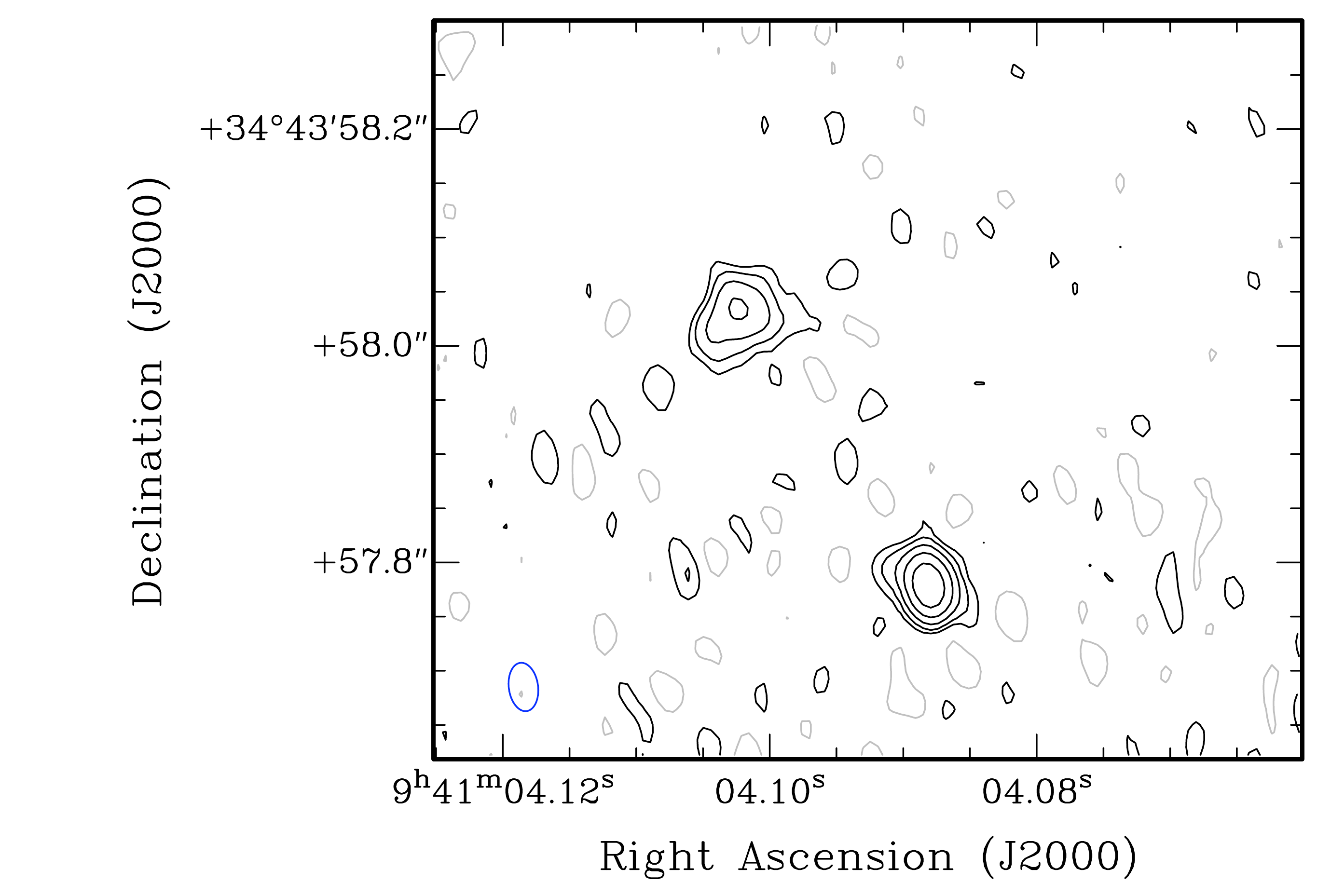}
\caption{e-VLBI $\lambda$ 18~cm radio map of IC~2497, showing both
  components, C1 $\&$ C2. The contours are at
  -1,1,2,4,8, and 16 $\times$ 0.03 ${\rm mJy}\,{\rm
    beam}^{-1}$.}

\end{figure}

\subsection{MERLIN 18~cm observations}
Additional observations of IC~2497 were conducted with MERLIN at
$\lambda$ 18~cm, using 15 x 1 MHz channels in all 4 Stokes
parameters. The observations were performed in two separate runs,
on the 2 February 2009 $\&$ 3 February 2009 (16:22 - 08:59 UT) for approximately 17
hours and on the 4 February 2009 $\&$ 5 February 2009 (17:17 - 07:20 UT), \fett{with approximately 15 hours and 12 hours on-source, respectively}.
The outer 2 frequency channels were deleted due to
band-pass effects. \fett{The UV coverage for this observation extends to a maximal baseline of 1.2 M$\lambda$. The shortest projected baseline at $\sim$ 90 k$\lambda$ corresponds to a spatial scale of $\sim$ 2.29 arcseconds, above which structures are resolved out.}

The observations were amplitude-calibrated using 3C~286 (the primary flux-density
calibrator) $\&$ B2134+004 (a point-source secondary amplitude
calibrator). We found B2134+004 to have a flux density of 10.058
Jy.  A phase reference cycle time of 8
minutes was employed (7 minutes on the target IC~2497, and 1 minute on
the phase calibrator, J0945+3534 \citep{VCS1}). The phase solutions
derived from J0945+3534 were transferred to the target, IC~2497. No
self-calibration techniques were applied to the target data, as the
source was very weak and heavily resolved on the Cambridge-Defford
baseline.  The data were re-weighted to reflect the relative telescope
sensitivities, in order to optimise the r.m.s. noise in the image. The
final naturally weighted image is presented in Fig. 2.

\subsection{MERLIN 6~cm observations}
We observed IC~2497 using MERLIN at $\lambda$ 6~cm, with 15 x 1 MHz
channels in all 4 Stokes parameters. The observations were performed in two
18.5 hour runs on 21 March 2009 $\&$ 22 March 2009 (12:15-07:00 UT) and 22 March 2009
$\&$ 23 March 2009 (12:30-07:00 UT), 
\fett{
with approximately 16 hours on-source. The maximum UV coverage for
this observation, extends to 3 M$\lambda$. The shortest projected
baseline covers $\sim$ 0.27 M$\lambda$, which corresponds to a spatial
scale of 0.76 arcseconds, above which structures are resolved out.} 
The calibration and analysis followed the same path as the 18~cm
observations. The final naturally weighted image, with an additional
2~M$\lambda$
Gaussian taper applied to the UV-data,
is shown in Fig. 3.

\begin{figure}

\centering 
\includegraphics[width=0.45\textwidth]{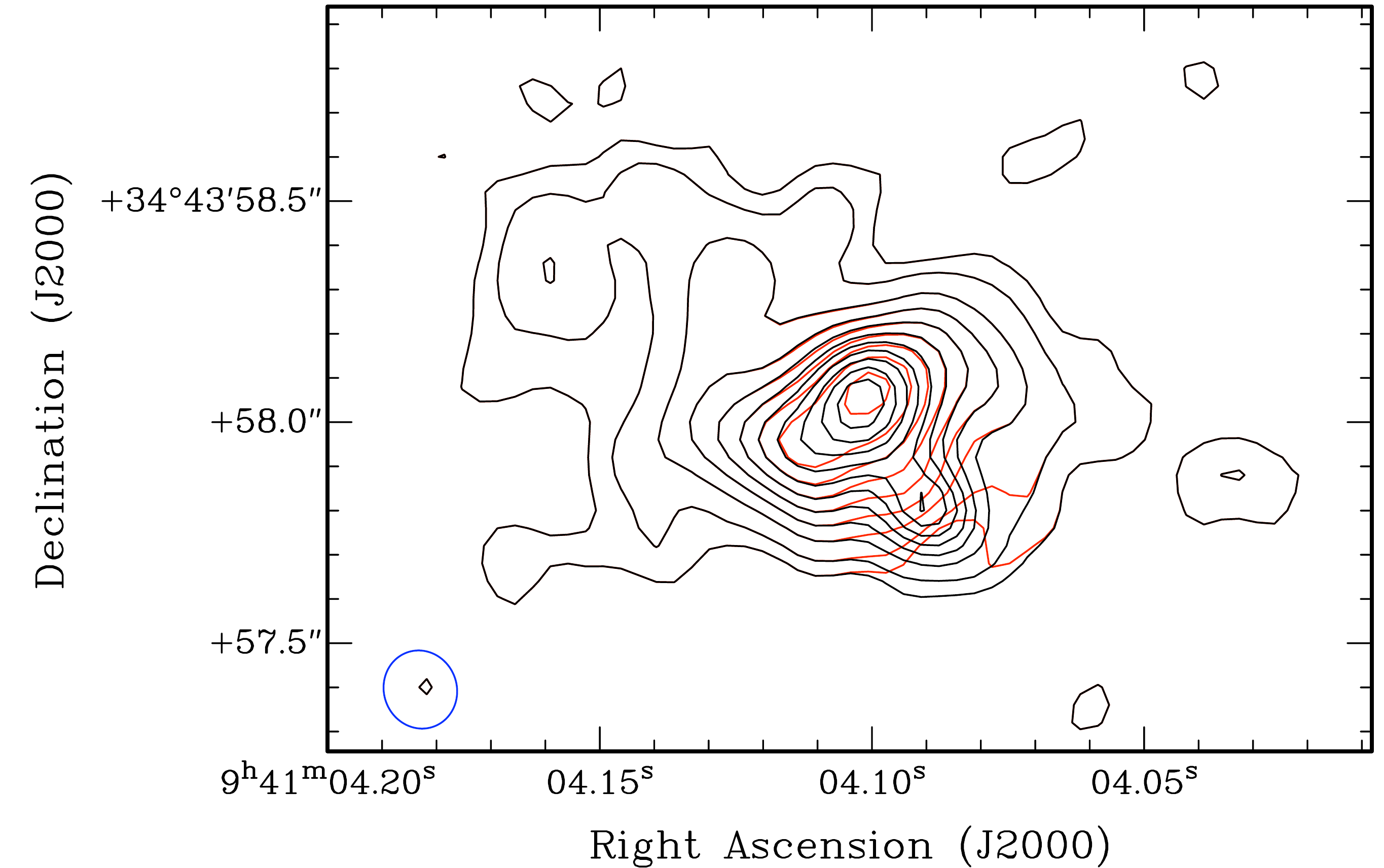}
\caption{
The naturally weighted $\lambda$ 18~cm MERLIN radio map of IC~2497
(black contours), showing both C1 $\&$ C2, embedded within a region of
smooth extended emission, overlaid over the same map with the point
sources subtracted (see text, grey contours, red in online
version). The contours are drawn at 1,2,4,6,8,10,12,14,16,18, and 20
$\times$ 0.074 ${\rm mJy}\,{\rm beam}^{-1}$.}
\end{figure}

\begin{figure}

\centering
\includegraphics[width=0.45\textwidth]{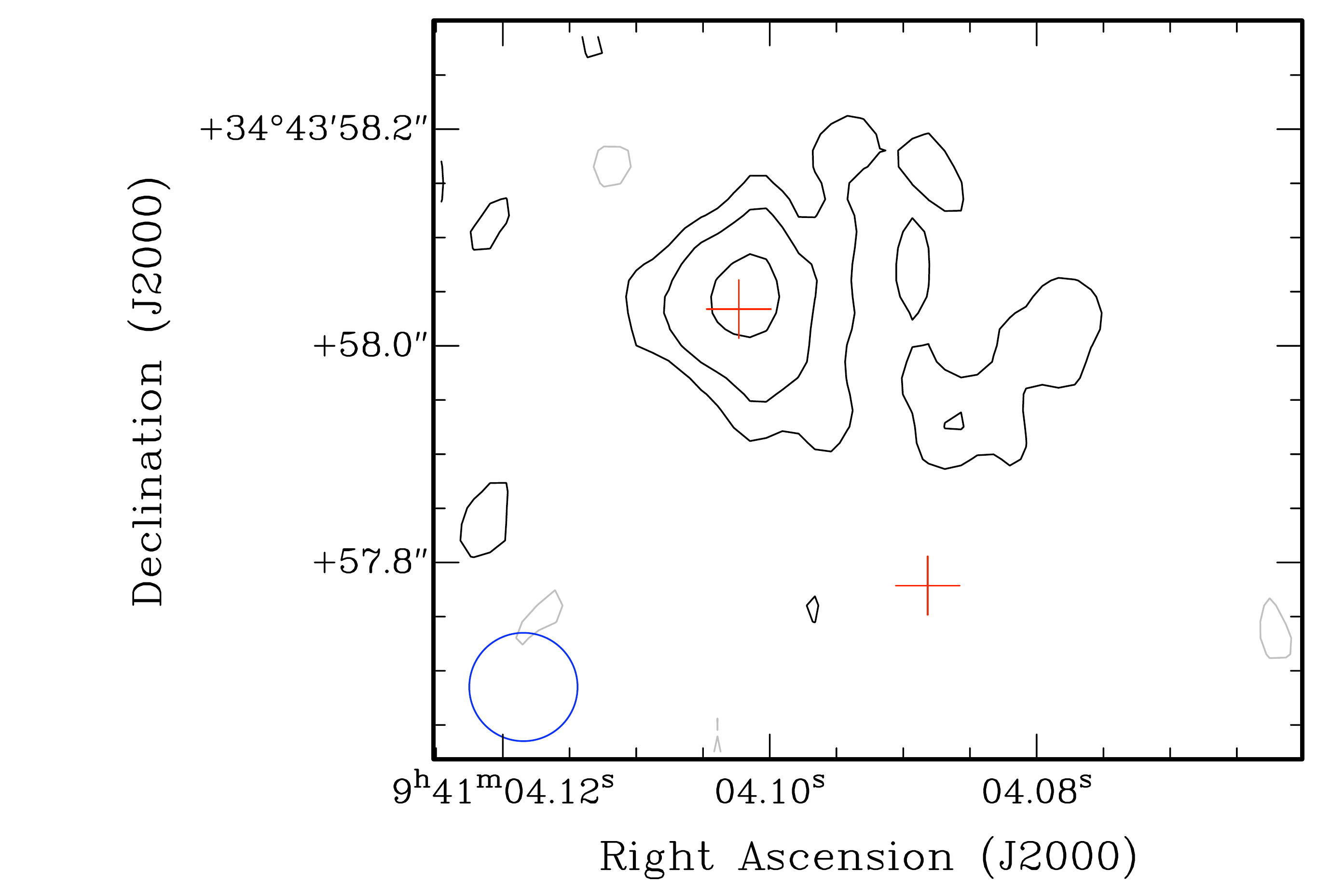} 
\caption{
$\lambda$ 6~cm MERLIN radio map of IC~2497, showing the component
C1. The contours are drawn at -1,1,2, and 4 $\times$ 0.140 ${\rm mJy}\,{\rm
beam}^{-1}$. The crosses indicate the positions of C1 and C2 as measured from the VLBI map.}
\end{figure}
\section{Results} 

Figure 1 shows the the radio image of the 1.65 GHz e-VLBI
observation. The image shows two compact radio components, separated
by $\sim$ 300 milliarcseconds ($\hat{\sim}$ 312 pc)\footnote{We use a
redshift-distance of 210 Mpc \citep{Joshetal09} throughout this
paper}. The component to the southwest (hereafter C1) was initially
detected by a short track 18~cm EVN (e-VLBI) observation
\citep{Joshetal09}, and to the northeast of C1 is a fainter, newly
discovered component (hereafter C2).

The sources appear marginally resolved in the VLBI map but this may be
due to residual phase errors after phase referencing. The total
integrated flux density measured for C1 is 1.03 $\pm$ 0.03 mJy,
\fett{which is close to the flux density of 1.1 $\pm$ 0.1 ${\rm mJy}$ 
obtained by \cite{Joshetal09}} The total integrated flux density for
C2 is 0.61
$\pm$ 0.04 mJy. This represents only 10$\%$ of the total WSRT and VLA
FIRST flux density measurements (20.9 and 16.8 mJy,
respectively). Measurements of the size of both components using the
AIPS task IMFIT suggest maximum sizes of 44 milliarcseconds for C1,
and 58 milliarcseconds for C2. From this, we determine a limit of the
brightness temperature of $T_{b} \,>\, \sim 4 \, \times \,10^{5}\,K$,
and $T_{b}\,>\,\sim 1.4\,\times\,10^{5}\,K$, for C1 and C2,
respectively.
 
In the lower resolution 1.65 GHz
MERLIN observation (see Fig. 2), C1 is also shown to be
compact, but with a possible slight elongation to the southwest,
while C2 shows a lot of
extended emission. The extended emission located around C2 has a
position angle that is similar to the large-scale major axis of the
disk in IC~2497.  We measure the peak brightness of C1 and C2 to be
$0.88\,{\rm mJy}\,{\rm beam}^{-1}$ and 1.00 ${\rm mJy}\,{\rm
beam}^{-1}$, respectively. The total flux density including both
components and the extended emission is 11.9 $\pm$ 0.2 mJy, which
corresponds to $ \approx 12.8\,{\rm mJy}$ at 1.4 GHz, using the
spectral index of -0.55, from \cite{Joshetal09}.
\fett{ Subtracting the flux density of the extended WSRT (large-scale jet-) component (S$_{1.4GHz} = 3.2 \pm\, 0.2\, {\rm mJy}$, \citealt{Joshetal09}), from the total WSRT flux density (S$_{1.4GHz} = 20.9 \pm\, 1.1\, {\rm mJy}$) leaves 17.7 ${\rm mJy}$ for the central point source as observed by the WSRT. Therefore, $\sim$ 75$\%$ of the emission from the VLA FIRST survey and $\sim$ 70$\%$ of the emission contained in the central, component, unresolved by the WSRT \citep{Joshetal09}, are recovered by the 18~cm MERLIN observation. This suggests that most of
the radio flux measured by the WSRT and VLA is associated with the extended emission detected in the MERLIN observations. To help determine of the structure of this extended component, the VLBI CLEAN components were convolved to match the MERLIN (18~cm) resolution, rescaled by a factor of 0.8 to minimise residuals, and subtracted from the MERLIN 18~cm map. The resulting residual map is shown in Fig.~2.}

We detect only component C2 in the MERLIN 5 GHz observations (see
Fig 3). None of the extended emission associated with C2 at 1.65
GHz is detected. At 5 GHz C2 appears to be slightly extended in the
direction of C1 at 18~cm. The component C1 is not detected, placing a 5-$\sigma$ upper
limit on its flux density of 0.26 mJy at this frequency and
resolution. The position of C1 detected by the EVN at 18~cm is
indicated by the cross in Fig. 3. A summary of the various
properties of components C1 and C2 are presented in Table 2.

\begin{table*}
\centering 
\caption{Results of EVN 18~cm and MERLIN 6 and 18~cm observations, for both C1 and C2.}
{\tiny
\addtolength{\tabcolsep}{-1pt}
\scalebox{0.7}{}
\begin{tabular}{ccccccccc}

\hline  
 Component & Observation & $\alpha_{J2000}$ & $\delta_{J2000}$ & Size & Physical Size & P.A. & $\textit{S}_{Peak}$ & $\textit{S}_{Total}$  \\ 
  &  &  &  & (mas $\times$ mas)& (pc $\times$ pc) & ($^{o}$)  & (${\rm mJy}\,{\rm beam}^{-1}$) & (${\rm mJy}$) \\
\hline
C1 & EVN 18~cm  & $09^{h} 41^{m} 04^{s}$.089 & +$34^{o}$ 43' 57".78 & 44.2$\pm$0.7 $\times$ 31.4$\pm$0.5 & 45.0 $\times$ 32.7 & 17$\pm$2 & 0.876$\pm$0.015 & 1.026$\pm$0.028 \\ 
   & MERLIN 18~cm  & $09^{h} 41^{m} 04^{s}$.090 & +$34^{o}$ 43' 57".82 & 222.5$\pm$10.6 $\times$ 160.3$\pm$7.6 & 231.4 $\times$ 166.7 & 35$\pm$5 & 1.076$\pm$0.035 & 2.464$\pm$0.110 \\ 
   & MERLIN  6~cm  & - & - & - & - & - & - \\ 
\hline
C2 & EVN 18~cm  & $09^{h} 41^{m} 04^{s}$.10 & +$34^{o}$ 43' 58".03 & 58.1$\pm$3.1 $\times$ 47.8$\pm$2.6 & 60.4 $\times$ 49.7 & 141$\pm$11 & 0.255$\pm$0.014 & 0.609$\pm$0.045 \\ 
   & MERLIN 18~cm  & $09^{h} 41^{m} 04^{s}$.10 & +$34^{o}$ 43' 58".02 & 453.3$\pm$17.8 $\times$ 282.4$\pm$11.1 & 471.4 $\times$ 293.7 & 125$\pm$3 & 1.430$\pm$0.034 & 9.007$\pm$0.2.43 \\ 
   & MERLIN  6~cm  & $09^{h} 41^{m} 04^{s}$.10 & +$34^{o}$ 43' 58".04 & 139.6$\pm$13.4 $\times$ 122.3$\pm$11.7 & 145.2 $\times$ 127.2 & 6$\pm$21 & 0.682$\pm$0.065 & 1.120$\pm$0.159 \\  
\hline 
\end{tabular} 
}
\end{table*}
\section{Discussion}
Our results support the hypothesis that an AGN is
located at the centre of IC~2497. In our 18~cm EVN and MERLIN
observations, we have detected 2 distinct radio sources in the central region of
IC~2497, (C1 and C2) with measured brightness temperatures in excess
of $10^{5}\,$K, \fett{an upper limit for brightness
temperatures of star-forming regions \citealt{CHYT91, Biggs10}}. By tapering the EVN image
($S_{C1\,total, \,2M\lambda\,taper} = 1.103\,{\rm mJy}\,\&~
S_{C2\,total,\,2M\lambda\,taper} = 0.597~mJy $) and comparing with the MERLIN 6~cm map, we derive a relatively flat spectral index ($S\propto \nu^{\alpha}$) $\alpha_{C2} \sim 0.12 \pm 0.01$ for C2, suggesting that this is the radio core in IC~2497
associated with a central AGN. By
comparison C1, has a much steeper spectrum: assuming an upper limit to
the flux density of C1 of $\sim \,0.26$~mJy, we derive a spectral
index of $\alpha_{C1}\, < \,-1.38\,\pm\,0.10$. In this scenario, C1 is
most likely a hotspot in the large-scale jet observed on larger
scales. Unfortunately, the radio luminosity of C2 tells us very little
about the ionising potential of the associated AGN at optical and UV
wavelengths. It does however, clearly identify some level of AGN
activity in the central regions of IC~2497. The approximate separation
of the components C1 and C2 is 300 milliarcseconds, with the associated
position angle being $215$ degrees. Given the different scales
involved, this is very similar to the position angle defined by both the
direction of the WSRT kpc jet \citep{Joshetal09} and the Voorwerp
itself of $\sim\, 280$ degrees. The extended emission detected by
MERLIN at 18~cm and associated with C2 is aligned with the major axis
of the galaxy. This extended emission has physical dimensions of $\sim \,0.4$ kpc.  A comparison of
the WSRT, MERLIN, and e-VLBI observations suggests that the bulk of the
radio emission is extended in nature.

\fett{ We can estimate the flux density of the radio emission associated with star formation  in IC2497 by subtracting the contribution from both the compact VLBI components and  the large-scale radio jet detected by the WSRT. This yields a flux density for the extended  radio emission associated with the star formation of 15.7 $\pm$ 1.1 mJy. This, in addition to the IRAS FIR flux densities of $S_{100\mu}$ = 3.04 $\pm$ 0.3 Jy and $S_{60\mu}$ = 1.66 $\pm$ 0.2 Jy \citep{IRAS}, permits us to  calculate a q-value  \citep{Condon92} of 2.2 $\pm$ 0.04. We find that IC2497 clearly,  lies close to the standard FIR-radio correlation for star-forming galaxies, and the sub-kpc  scale of the extended emission implies that it is a good example of a nuclear  starburst system. It appears that this nuclear starburst is coincident with the AGN core. The luminosity of the extended radio emission implies a massive star formation rate. Using the standard relations given in Condon (\citeyear{Condon92, Condon02}) with the implied standard IMFs \citep{MS79,Salpeter55}, we derive a star formation rate ${\rm SFR}(M > 5 \,M_{\odot}$) of $18.1 \,M_{\odot}\,{\rm yr}^{-1}$ (or ${\rm SFR}(M\, >\, 0.1\, M_{\odot}) \sim 100 \,M_{\odot}\,{\rm yr}^{-1}$) for the entire galaxy. The star formation rate derived from the central, extended emission detected by MERLIN is ${\rm SFR}(M\, >\, 5\, M_{\odot}) \sim 12.4 \,M_{\odot}\,{\rm yr}^{-1}$ (or ${\rm SFR}(M > 0.1 \,M_{\odot})\sim 68 \,M_{\odot}\,{\rm yr}^{-1}$).}

\fett{With a FIR luminosity of $1.3\,\times\,10^{11}{\rm L}_\odot$ derived from the IRAS flux densities, IC~2497
lies in the regime of a class of galaxies known as the luminous infrared galaxies (LIRGS). It also
shares their characteristics of a nuclear starburst
\citep{Sanders96}, as well as showing clear signs of interaction with
the environment \citep[][see also below]{Joshetal09}. This nuclear
starburst is likely to be responsible for significant obscuration of the
central AGN along our line-of-sight \citep{Sanders96}. The
enhanced level of star formation in IC~2497 is also presumably associated with
the infall of gas due to previous interactions with neighbouring
galaxies in the field, which is assumed in general to be the origin of
the enhanced star formation in LIRGS \citep{Sanders96}. This
interaction would induce large-scale non-circular motions that could
transport gas to the central regions, fueling both the nuclear
starburst and AGN activity simultaneously in IC~2497.}

Another consequence of past-interactions, is the creation of
a huge debris reservoir of HI gas around the galaxy, as observed by
the WSRT. We propose that Hanny's Voorwerp corresponds to the part of
the debris trail that is illuminated by the AGN illumination cone (and
indeed the nuclear starburst). The line-of-sight between Hanny's
Voorwerp and the nuclear region of IC~2497 is roughly perpendicular
to the plane of the major axis of IC2497, and is therefore likely to
be relatively unobscured.

In this scenario, we note that another possibility is that
components C1 and C2 can be identified with luminous
SNe or SNR. This applies in particular to C1, which may exhibit a high
level of variability \fett{as indicated by its non-detection in the original e-VLBI observations \citep{Joshetal09}.} 
However, we believe this to be very unlikely
given the flat spectrum of C2 and the implied luminosities of the SNe,
which would have to be in excess of $5.126\,\times\,10^{21}$
Watts/Hz i.e.  almost an order of magnitude brighter than the
supernovae observed in Arp 220 \citep{Parraetal07}. \fett{In
addition, C1 appears to have a constant flux density over a time span
of $\sim 8$ months, making a supernova interpretation extremely
unlikely for this source.}

\fett{ If AGN activity is responsible for the ionisation of Hanny's Voorwerp,
the AGN must have been radio quiet even at the time at which the
ionising radiation was generated. At 1.4 GHz, assuming
a typical magnetic field strength of $\sim 100 \mu$G, the characteristic life
time of the relativistic jet electrons is
$\sim\,9\,\times\,10^{6}$ years (e.g. \citealt{Condon92}), exceeding the
light travel time to Hanny's Voorwerp by a factor of ~10. Hence, to
judge whether the AGN at the centre of IC 2497 could have reached the
bolometric or ionising luminosity sufficient to ionise Hanny's
Voorwerp, one would need to search for radio-quiet AGN with similar
ionisation structures at large distances. However, while targeted
surveys of very luminous radio galaxies have previously discovered highly ionised gas \citep{Tadhunter02, Tadhunter07} at large
distances ($>$ 10 kpc) from AGN, little is known about AGN with low
radio luminosities such as that observed in IC 2497.} 

\section{Summary}
In conclusion, our results suggest that IC~2497 is an "active" galaxy
in the broadest possible sense - it exhibits evidence of both a \fett{highly obscured} AGN and
nuclear star-forming activity in its central regions. There is also
evidence (e.g. HI absorption, see \citealt{Joshetal09}) that our
line-of-sight towards the nuclear regions of IC~2497 is obscured. Past
interactions between IC~2497 and neighbouring galaxies have presumably
triggered and fuelled both the AGN and enhanced star-forming activity
observed, and are likely to be responsible for the creation of a huge
reservoir of HI gas around the galaxy.  We believe that Hanny's
Voorwerp corresponds to that part of the debris trail that is
illuminated by the AGN ionisation cone (and indeed the nuclear starburst).

Many AGN with weak
radio sources exist but the presence of a large surrounding gas
reservoir is most likely only present in interacting systems \fett{(such as LIRGs, which IC 2497 seems to be a prototype of)}. 
Given that galaxy interactions are relatively uncommon in the nearby universe, phenomena
such as Hanny's Voorwerp while appearing extremely dramatic are expected to be
quite rare. In the case of the Voorwerp, it seems that the location of
this gas reservoir is also important, i.e., it must lie in the vicinity
of the unobscured AGN illuminating cone.


\begin{acknowledgements}
This research and HR was supported by the EC Framework 6 Marie Curie
Early Stage Training programme under contract number
MEST-CT-2005-19669 "ESTRELA". Support for the work of KS was provided
by NASA through Einstein Postdoctoral Fellowship grant number
PF9-00069 issued by the Chandra X-ray Observatory Center, which is
operated by the Smithsonian Astrophysical Observatory for and on
behalf of NASA under contract NAS8-03060. The European VLBI Network is
a joint facility of European, Chinese, South African and other radio
astronomy institutes funded by their national research councils.
e-VLBI developments in Europe are supported by the EC DG-INFSO funded
Communication Network Developments project 'EXPReS', Contract
No. 02662. MERLIN is a National Facility operated by the University of
Manchester at Jodrell Bank Observatory on behalf of STFC.

\end{acknowledgements}

\bibliographystyle{aa}
\bibliography{14782}

\end{document}